\documentclass[a4paper,12pt]{article}
\usepackage[dvips]{graphicx}
\usepackage{amsmath,amssymb}

\begin{document}
%===============================================================
\title{Broken $S_3$ Symmetry in Flavor Physics}
\author{Toshiaki Kaneko${}^{(a)}$ and Hirotaka Sugawara${}^{(b)}$
\\
\\
{\small \textsl{$(a)$ Computing Research Center},}
\\
{\small \textsl{High Energy Accelerator Research Organization (KEK),}}
\\
{\small \textsl{Tsukuba, Japan}}
\\
{\small \textsl{$(b)$ JSPS, Japan Society for Promotion of Science,}}
\\
{\small \textsl{Washington Office, Washington, DC, USA} }
}

% \date{(Version 2010/11/18)}
\date{}
\maketitle

\begin{abstract}
The $S_3$ symmetry is shown to be a very good approximate
symmetry when it is broken in a specific way.
This is true both in quark sector and in lepton sector.
The way to break it is implied by the K-M mechanism applied not to the
mixing matrix but to the mass matrices.
In quark sector, we have an almost perfect fitting to the experimental
data, and in lepton sector, we have a precision for the $\theta_{13}$.
\end{abstract}

\newpage
%===============================================================

Some time ago, one of the authors (H.S.) together with S.~Pakvasa made a
proposal to understand the flavor physics within the framework of
$S_3$ symmetry\cite{cite01}.
Recently, we see some revival of this idea\cite{cite02} together with
possibilities of other discrete symmetry\cite{cite03}.
In this letter we show that $S_3$ group is a very good symmetry if it is
broken in a specific way.
This is true both in quark and in lepton sectors.
We have a perfect fit to the CKM matrix in quark sector and some
predictions in lepton sector: We have a one parameter description of the
neutrino mixing matrix which becomes the tri-bimaximal\cite{cite04} 
in the limit of vanishing $\theta_{13}$.
In fact, if we take the central value of the KamLAND data\cite{kamland} on
$\theta_{12}$, we predict that the value of $\sin^2\theta_{13}$ must be
somewhere around 0.02.

It is well known by now that the three generation mass matrix (whether
for quark or lepton sector) must have the following form if we assume
the $S_3$ symmetry, the $S_3$ double-singlet quarks or leptons and
the $S_3$ singlet Higgs:
\begin{equation}
M = 
\left(\begin{array}{ccc} \label{s3}
  a & b & b \\
  b & a & b \\
  b & b & a
\end{array}\right)
.
\end{equation}
This cannot be exact because, for example, it predicts at least
two of the masses to be degenerate which is not true.
The question is, therefore, what the proper way is to violate the $S_3$
symmetry.
The clue is given by the K-M mechanism of CP violation.
K-M mechanism\cite{cite06} is to violate the CP by giving all possible 
phases to the quark mixing matrix.
Since the mixing matrix is given by diagonalizing the mass matrices, we
argue that the K-M mechanism should be applied to the more fundamental
mass matrices rather than mixing matrix.
This implies that we use, rather than (\ref{s3}), the following mass
matrix:
\begin{equation} \label{s3delta}
M = 
\left(\begin{array}{ccc}
  a e^{i\delta_{11}} & b e^{i\delta_{12}} & b e^{i\delta_{13}} \\
  b e^{i\delta_{21}} & a e^{i\delta_{22}} & b e^{i\delta_{23}} \\
  b e^{i\delta_{31}} & b e^{i\delta_{32}} & a e^{i\delta_{33}}
\end{array}\right)
.
\end{equation}
Five of these phases can be absorbed in the wave functions leaving four
phases and $a, b$ as independent parameters.
In the following we discuss the consequence of equation (\ref{s3delta})
both in the quark and the lepton sector.

\vspace{1em}

\noindent
(A) Quark sector

Starting from equation (\ref{s3delta}), we construct the $M M^\dagger$ 
for which we get the following :
\begin{equation} \label{fgh}
M M^\dagger = 
\left(\begin{array}{ccc}
  k            & f & \overline{g} \\
  \overline{f} & k & h \\
  g            & \overline{h} & k
\end{array}\right)
.
\end{equation}
with
\begin{equation} \label{cond:fgh}
k \text{: real}
,\quad
\text{Re} f \leq k
,\quad
\text{Re} g \leq k
,\quad
\text{Re} h \leq k
\end{equation}
We can show that the form (\ref{fgh}) with the condition (\ref{cond:fgh}) 
is equivalent to (\ref{s3delta}).
In this case, only one phase can be absorbed in the wave function
because the right handed quark wave function does not appear when we
take the matrix element of (\ref{fgh}).
This leaves us 6 parameters as before.
In fact we can absorb one more phase to the left handed up/down
quark wave function but, by so doing
we will not be able to absorb any phase from down/up quark wave function
respectively, leaving always the 12 parameters.
In the following we use the convention in which $f$ and $h$ of either up
or down quark $M M^\dagger$ are real.

The relation of the parameters to the quark masses is the following:
\begin{align}
k &= \frac{1}{3} s_1
,\\
|f|^2 + |g|^2 + |h|^2 &= \frac{1}{3} s_1^2 - s_2
,\\
f g h + \overline{f}\overline{g}\overline{h}
 &= \frac{2}{27} s_1^3 - \frac{1}{3} s_1 s_2 + s_3
\end{align}
with
\begin{align*}
s_1 = m_1^2 + m_2^2 + m_3^2
,\quad
s_2 = m_1^2 m_2^2 + m_2^2 m_3^2 + m_3^2 m_1^2
,\quad
s_3 = m_1^2 m_2^2 m_3^2
.
\end{align*}
We use 6 quark masses\cite{pdg} 
to reduce the number of parameters from 12 to 6.
The remaining 6 parameters are fitted to quark mixing matrix (CKM matrix)
which is written by 4 real parameters.
The result of the fitting of the matrix $M M^\dagger$ is given below.
The convention we adopt is the reality of the parameters $f$ and $h$ 
of $M M^\dagger$ of up quarks.%
\footnote{
More than 10 digits should be taken into account in order to reproduce
quark mass spectrum, since their hierarchy structure is represented
by small differences among values of order one.
}
\begin{align*}
(M M^\dagger)_u = k_u
  \left(\begin{array}{ccc}
   1.000  &  0.9999  &  
   \left\{\begin{array}{c} 0.9999 \\ -0.00001431 i \end{array}\right\}
   \\
   0.9999 &  1.000   & 1.0000  
   \\
   \left\{\begin{array}{c} 0.9999 \\ + 0.00001431 i \end{array}\right\} & 
   1.0000 &  1.000 
  \end{array}\right)
\end{align*}
\begin{align*}
(M M^\dagger)_d = k_d
  \left(\begin{array}{ccc}
   1.000  &  
   \left\{\begin{array}{c} 0.9955 \\ +0.08036 i \end{array}\right\} &
   \left\{\begin{array}{c} 0.9944 \\ +0.09255 i \end{array}\right\}
   \\
   \left\{\begin{array}{c} 0.9955 \\ -0.08036 i \end{array}\right\} &
   1.000 &
   \left\{\begin{array}{c} 0.9999 \\ +0.01278 i \end{array}\right\}
   \\
   \left\{\begin{array}{c} 0.9944 \\ -0.09255 i \end{array}\right\} &
   \left\{\begin{array}{c} 0.9999 \\ -0.01278 i \end{array}\right\} &
   1.000  
  \end{array}\right)
\end{align*}

These matrices give the following CKM matrix:
\begin{equation*}
(\text{CKM}) =
  \left(\begin{array}{ccc}
     0.9743 &  
     0.2253 &  
     \left\{\begin{array}{c} 0.001249 \\ -0.003232 i \end{array}\right\}
     \\
     \left\{\begin{array}{c} -0.2251 \\  -0.000136 i \end{array}\right\}
     &
     0.9735 &  
     0.04102 \\
     \left\{\begin{array}{c} 0.008026 \\ -0.003146 i  \end{array}\right\} &
     \left\{\begin{array}{c} -0.04025 \\ -0.0007286 i \end{array}\right\} &
     0.9992 
  \end{array}\right)
\end{equation*}
Equivalently the Wolfenstein parameters\cite{cite07} are given in 
table~\ref{tab:wparam}.

{\small
\begin{table}[!ht]
\begin{center}
\begin{tabular}{cll}
\hline
W-Parameters & Experimental & Calculated \\
\hline
$\lambda$ & $0.2253 \pm 0.0007$ & 0.225267 \\
$A$       & 
    $0.808  {\tiny \begin{array}{l} +0.022 \\ -0.015
                   \end{array}}$ 
          & 0.808417 \\
$\bar{\rho}$ & $0.132  {\tiny \begin{array}{l} +0.022 \\ -0.014 
                         \end{array}}$
          & 0.131926 \\
$\bar{\eta}$ & $0.341 \pm 0.013$ & 0.340840 \\
\hline
\end{tabular}
\end{center}
\caption{Fitted Wolfenstein parameters.
    \label{tab:wparam}}
\end{table}
}

Our convention for the CKM matrix is not the usual one but the
perfection of our fitting
is clear from the table of the convention-independent Wolfenstein
parameters.
The important point is
that our mass matrices for up and down quarks 
($M M^\dagger$ matrices to be precise) are very close to the
$S_3$ invariant ones indicating that the breaking of $S_3$ is small.
In fact the imaginary part of $f$, $h$ or $g$ is always less than 10 \%
of the real part. 
It is also very impressive that
the off-diagonal elements are all smaller than the diagonal element as
the condition (\ref{cond:fgh}) indicates but the deviation
is only less than 0.1 \%.
This corresponds to the smallness of $2 \otimes 2$ mass 
of $(2+1)\otimes(2+1)$ in $S_3$
implying the approximate democracy.

\vspace{1em}

\noindent
(B) Lepton sector

Lepton sector is more complicated than the quark sector due to the
structure of the neutrino mass matrix. 
We adopt the generalized version of the see-saw mechanism\cite{cite08} in the
following way :
\begin{equation}
M_\nu^{(6)} = (\nu_L^T, (\nu_R^C)^T) C
  \left(\begin{array}{cc}
     0 & M^D \\ (M^D)^T & M^M 
  \end{array}\right)
  \left(\begin{array}{c}
     \nu_L \\ \nu_R^C
  \end{array}\right)
,
\end{equation}
where
$M^M = (M^M)^T$ is a $3 \times 3$ right handed Majorana mass matrix.
The $6\times 6$ eigenvalue equation becomes,
\begin{equation}
  \left(\begin{array}{cc}
     0 & M^D \\ (M^D)^T & M^M 
  \end{array}\right)
  \left(\begin{array}{c}
     V_1 \\ V_2
  \end{array}\right)
= \lambda
  \left(\begin{array}{c}
     V_1 \\ V_2
  \end{array}\right)
\end{equation}
For small $\lambda$ (compared with the large eigenvalues of 
$M^M$ which we assume),
we get,
\begin{equation}
- M^D (M^M)^{-1} (M^D)^T V_1 =
 \lambda V_1 \quad\text{ and }\quad V_2 \sim 0
.
\end{equation}
This implies the following extended see-saw equation,
\begin{equation} \label{mnu3}
M_\nu^{(3)} = - M^D (M^M)^{-1} (M^D)^T
.
\end{equation}
We assume that both $M^D$ and $M^M$ have the form of equation
(\ref{s3delta}).
Taking into account the fact that $M^M$ is a complex symmetric matrix and
also assuming that the $S_3$ breaking comes
only from the phases of $M^M$, we write.
\begin{align} \label{mdmat}
M^D &= m_\nu
\left(\begin{array}{ccc}
  1 & c & c \\
  c & 1 & c \\
  c & c & 1
\end{array}\right)
,\\ \label{mmmat}
M^M &= a
\left(\begin{array}{ccc}
  1                  & \eta e^{i\delta_1} & \eta e^{i\delta_2} \\
  \eta e^{i\delta_1} & 1                  & \eta e^{i\delta_3} \\
  \eta e^{i\delta_2} & \eta e^{i\delta_3} & 1
\end{array}\right)
.
\end{align}
Substituting equations (\ref{mdmat}) and (\ref{mmmat}) 
into equation (\ref{mnu3}) we get,
\begin{equation} \label{mnumat}
\begin{split}
M_\nu^{(3)} = \frac{m_\nu^2 \epsilon^2}{a z}
   \Bigl[ &  \chi
\left(\begin{array}{ccc}
  -2 &  1 & 1 \\
   1 & -2 & 1 \\
   1 &  1 & -2
\end{array}\right)
\\
&
- i
\left(\begin{array}{ccc}
  -(x_1+x_2) &  x_1       & x_2       \\
   x_1       & -(x_3+x_1) & x_3       \\
   x_2       &  x_3       & -(x_2+x_3)
\end{array}\right)
\Bigr]
,
\end{split}
\end{equation}
where
\begin{gather*}
\epsilon = c - 1
,\quad 
\chi = 1 - \eta
,\quad
z = 1 + q_1 q_2 q_3 - q_1^2 - q_2^2 - q_3^2
,\\
q_j = \eta e^{i \delta_j}
,\quad
x_j = \delta_{j+1} + \delta_{j+2} - \delta_{j}
.
\end{gather*}
$M_\nu^{(3)}$ is generically a complex symmetric matrix and, 
according to the Takagi's factorization theorem\cite{cite09},
it can be diagonalized in the following way,
\begin{align*}
M_\nu^{(3)} = U_\nu^T \Sigma U_\nu
,
\end{align*}
where $U_\nu$ is a unitary matrix. 
In our particular case of equation (\ref{mnumat}), $U_\nu$
becomes an orthogonal matrix
due to the fact that the real and the imaginary parts of $M_\nu^{(3)}$
commute with each other.
We must distinguish two cases :
\begin{itemize}
\item[(1)] 
Normal hierarchy case : $x_1 x_2 + x_2 x_3 + x_3 x_1 = 0$
\begin{align} \label{unumat}
U_\nu =
\left(
\frac{1}{N_1}
\left(\begin{array}{c}
   x_2 + 2 x_3 \\ - (2 x_2 + x_3) \\ x_2 - x_3
\end{array}\right)
,
\frac{1}{\sqrt{3}}
\left(\begin{array}{c}
   1 \\ 1 \\ 1
\end{array}\right)
,
\frac{1}{N_3}
\left(\begin{array}{c}
   x_2 \\ x_3 \\ - (x_2 + x_3)
\end{array}\right)
\right)
\end{align}
The first column corresponds to an eigenvalue 
% $3 \chi$, 
$ - 3 m_\nu^2 \epsilon^2 \chi/(a z)$,
the second column to eigenvalue zero and the third one to 
$ - m_\nu^2 \epsilon^2 \{ 3 - 2 i(x_1+x_2+x_3) \}\chi/(a z)$.
We see that $x_2 = 0$ in (\ref{unumat}) reduces exactly to tri-bimaximal
case.
Therefore, $x_2$ is the amount of the deviation of  $U_\nu$
from the tri-bimaximal matrix.

\item[(2)] 
Inverted hierarchy: $x_1 = x_2 = x_3$
\begin{align}
U_\nu =
\left(\begin{array}{ccc}
  \frac{1}{\sqrt{6}}   &   \frac{1}{\sqrt{2}} & \frac{1}{\sqrt{3}} \\
  \frac{1}{\sqrt{6}}   & - \frac{1}{\sqrt{2}} & \frac{1}{\sqrt{3}} \\
  - \sqrt{\frac{2}{3}} &  0                   & \frac{1}{\sqrt{3}}
\end{array}\right)
\end{align}
% The first and the second columns correspond to eigenvalue 
% $3 \chi -3 i x_j$
The first column corresponds to eigenvalue 
$ - m_\nu^2 \epsilon^2 (3 \chi - 3 i x_2)/(az)$,
the second column to 
$ - m_\nu^2 \epsilon^2 \{3 \chi - i (2 x_1 + x_2)\}/(az)$
and the third column to zero eigenvalue.
This does not correspond to the tri-bimaximal solution.
\end{itemize}

A few comments are in order regarding our results :

(1) There may be a correction from the phases of $M^D$ which we assume to be
vanishing, namely,
the $S_3$ is violated only in the phases of $M^M$.

(2) The mixing matrix is actually $U_\nu \times U_l$ 
where $U_l$ is a matrix diagonalizing the
charged
lepton mass matrix.
The fact that $U_\nu$ alone gives the experimentally correct
answer implies that
$U_l$ must be identity and $M_l$ must be diagonal from the beginning.
% The only way to reconcile this
% with the $S_3$ symmetry is to assume that all three charged leptons belong
% to $S_3$ singlet coupled to
% the singlet Higgs field independently.
This means that the $S_3$ breaking in the charged lepton sector
is not by the phase factor.
This gives one important clue in
constructing a model for flavor physics.
One might think that there may be a solution in which neutrino mixing
matrix is identity and the charged
lepton mixing matrix gives the tri-bimaximal solution. 
We checked that
this is not the case.

\begin{figure}[!ht]
    \begin{center}
    \includegraphics[width=10cm,clip=true]{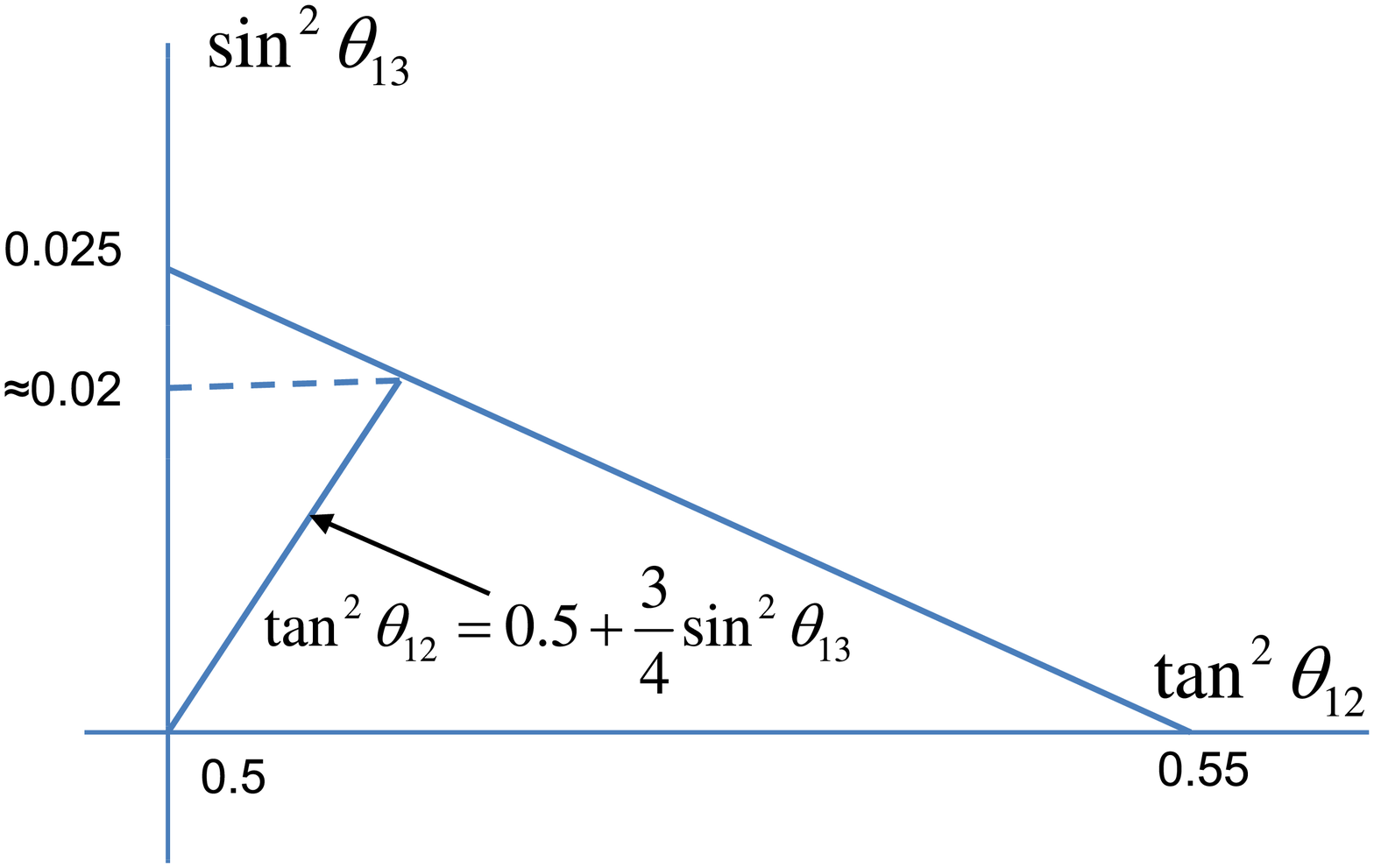}
    \caption{The value of $\sin^2\theta_{13}$.
             KamLAND data is taken from Fig.5 of Ref. \cite{cite05}.
             \label{fig:theta13} }
    \end{center}
\end{figure}

Next, we note that the mixing matrix given by equation (\ref{unumat}) is a one
parameter ($x_2/ x_3$) description of
the neutrino mixing matrix.
The usual $\theta_{12}$, $\theta_{23}$ and $\theta_{13}$ 
are all written by one parameter.
We write this in the following form in the case of small $\theta_{13}$ :
\begin{align} \label{sin13}
\sin^2 \theta_{13} &= \kappa
,\\ \label{tan12}
\tan^2 \theta_{12} &= 0.5 + \frac{3}{4} \kappa
,\\ \label{tan23}
\tan^2 \theta_{23} &= 1 \pm 2 \sqrt{2 \kappa}
.
\end{align}
Since we already have experimental values for $\theta_{12}$ and
$\theta_{23}$
we should be able to
predict the value of $\theta_{13}$.  
Here we use the KamLAND data for the $\theta_{12}$ and
equation (\ref{sin13}) and (\ref{tan12}) as an example to obtain $\theta_{13}$.
The fig.~\ref{fig:theta13} shows that
the value of $\sin^2\theta_{13}$ is approximately $0.02$.
This is when we take the central
value of KamLAND data\cite{kamland} for $\tan^2\theta_{12}$. 
We note that the central value of $\tan^2\theta_{12}$ 
for the SuperK data\cite{cite10} is less than $0.5$ 
which is in contradiction to
equation (\ref{tan12}) combined with (\ref{sin13})%
\footnote{
The most recent data of KamLAND\cite{kamland} implies
that $\tan^2\theta_{12} > 0.5$ for the reactor neutrino
but $\tan^2\theta_{12} < 0.5$ when it is combined with the solar neutrino
data.
}
.

{\small
\begin{table}[!ht]
    \begin{center}
\begin{tabular}{|cc|cc|cc|}
\hline
  $\sin^2\theta_{12}$ & $\tan^2\theta_{12}$ & 
  $\sin^2\theta_{23}$ %
                      & $\tan^2\theta_{23}$ & 
  $\sin^2\theta_{13}$ & $\tan^2\theta_{13}$ \\
\hline
   0.3350 & 0.5038 & 0.5498 & 1.2213 & 0.0050 & 0.0050 \\
   0.3367 & 0.5076 & 0.5702 & 1.3266 & 0.0100 & 0.0101 \\
   0.3384 & 0.5115 & 0.5856 & 1.4134 & 0.0150 & 0.0152 \\
   0.3401 & 0.5155 & 0.5985 & 1.4908 & 0.0200 & 0.0204 \\
   0.3419 & 0.5195 & 0.6098 & 1.5625 & 0.0250 & 0.0256 \\
   0.3436 & 0.5236 & 0.6198 & 1.6302 & 0.0300 & 0.0309 \\
   0.3454 & 0.5277 & 0.6289 & 1.6949 & 0.0350 & 0.0363 \\
   0.3472 & 0.5319 & 0.6373 & 1.7574 & 0.0400 & 0.0417 \\
   0.3490 & 0.5362 & 0.6451 & 1.8181 & 0.0450 & 0.0471 \\
   0.3509 & 0.5405 & 0.6525 & 1.8774 & 0.0500 & 0.0526 \\
\hline
\end{tabular}
    \caption{Mixing parameters for various values of
             $\kappa (= \sin^2 \theta_{13})$ from
             0.005 to 0.05.
             ({This corresponds to the positive sign
               in equation (\ref{tan23}) which in turn
               implies $\tan \theta_{12} \sin\theta_{13} > 0$.
             })
             \label{talbe:theta}}%
    \end{center}
\end{table}
}

Finally, we make some brief comments
on the model building based on our phenomenological analysis.  

(1) The
fact that we seem to have several independent phases imply that we may
have more than one Higgs fields which provide appropriate phases. 
We
cannot, of course, deny that all the phases are connected to a single
phase but it is not practical to construct such a model.  

(2) The
charged leptons belong to the 3 independent $S_3$ singlet whereas all the
others including the right handed neutrinos belong to the 
singlet-doublet.
The model must be able to explain this phenomenon.  

(3) $M^M$ and $M^D$ are
symmetric matrices. We checked that quark mass matrices are also
complex symmetric matrices up to the accuracy of better that 0.6\%.
This
suggests that the Higgs coupling to quarks or leptons must have the form
$\overline{\Psi}_i \Psi_j H^{ij}$
where $H^{ij}$ is symmetric in $i$ and $j \;(i, j = 1,2,3)$. 
When the vacuum value $\langle H^{ij} \rangle$ is such
that it satisfies 
$\langle H^{11} \rangle = \langle H^{22} \rangle = \langle H^{33} \rangle$
and $\langle H^{ij} \rangle = v\; (i \neq j)$, 
we get the $S_3$ invariant mass matrix,
if the vacuum values are all real.
The $S_3$ breaking is provided by the
phases of these vacuum values.
A model can be easily and naturally constructed if we combine these
phenomenological observations
with the following theoretical argument:
All the discrete symmetries and
some global symmetries are
an artifact of more fundamental gauge symmetry.
Namely, they arise when
some gauge symmetry is
broken at some high energy.
P, CP, R, Baryon number and lepton number all
belong to this category\cite{cite11}
and so is $S_3$.
These global symmetries are valid below the energy of the
breaking of original gauge symmetry.
The simplest gauge symmetry we can think of in our case is
$SU_3$.
We remind the reader that
we get this group when we break $E_8$ to $E_6 \times SU_3$.
This $SU_3$ can be shown to be infrared free\cite{cite12} and
its coupling may be very small at low energy but it could play a very
important role in flavor physics.
We hope to report the result of such a model in a future publications.

%===============================================================

%===============================================================

\end{document}